\documentclass[12pt]{article}
\usepackage[dvips]{graphicx}
\usepackage{latexsym}

\newcommand{\N}		{{\cal N}}
\newcommand{\eff}	{{\rm eff}}
\newcommand{\EW}	{{\rm EW}}
\newcommand{\BH}	{{\rm BH}}
\newcommand{\planck}	{{\rm pl}}
\newcommand{\DW}	{{\rm DW}}
\newcommand{\mini}	{{\rm min}}
\newcommand{\GUT}	{{\rm GUT}}

\newcommand{\CP}	{{\rm CP}}
\newcommand{\focus}	{{\rm fcs}}
\newcommand{\lnear}	{{\begin{array}{c} < \\[-0.8em] \sim \end{array}}}

\newcommand{\GeV}	{{\rm \;GeV}}

\newcommand{\fig}[1]	{Figure \ref{#1}}
\newcommand{\eq}[1]	{equation (\ref{#1})}

\begin{document}

\begin{flushright}
 \begin{minipage}[b]{43mm}
  YITP-02-14\\
  hep-th/0203010\\
  March 2002
 \end{minipage}
\end{flushright}

\renewcommand{\thefootnote}{\fnsymbol{footnote}}
\begin{center}
 {\large\bf
 Dynamical Formation of Spherical Domain Wall\\[3mm]
 by Hawking Radiation and\\[5mm]
 Spontaneous Charging-up of Black Hole}\\
 \vspace*{3em}
 {Yukinori Nagatani}\footnote
 {e-mail: nagatani@yukawa.kyoto-u.ac.jp}\\[1.5em]
 {\it Yukawa Institute for Theoretical Physics, Kyoto University,\\
      Sakyo-ku, Kyoto 606-8502, Japan}
\end{center}
\vspace*{1em}

\begin{abstract}
 We discuss the Hawking radiation in the Higgs-Yukawa system
 and we show dynamical formation of a spherical domain wall
 around the black hole.
 The formation of the spherical wall is a general property of the black hole
 whose Hawking temperature is equal to or greater than
 the energy scale of the system.
 The formation of the electroweak wall
 and that of the GUT wall
 are shown as realistic cases.
 We also discuss
 a phenomenon of the spontaneous charging-up of the black hole.
 The Hawking radiation can charge-up the black hole itself
 by the charge-transportation due to the spherical wall
 when C- and CP-violation in the wall are assumed.
 The black hole with the electroweak wall can obtain
 a large amount of the hyper charge.
\end{abstract}

%

\newpage
\section{INTRODUCTION}\label{intro.sec}

The radiation from a black hole known as the Hawking radiation
has a thermal spectrum \cite{Hawking:1975sw}.
In the case of a Schwarzschild black hole,
its temperature, namely, the Hawking temperature
is inversely proportional to mass of the black hole.
The black hole is losing its mass by the Hawking radiation
and it is increasing its Hawking temperature.
At the final stage of the radiation,
its temperature and its intensity explosively increase,
therefore the final stage of the black hole is very interesting
to particle physicists \cite{Hawking:1974rv}.

Several authors discussed that
the heating-up by the Hawking radiation can 
thermalize the neighborhood of the black hole
and thermal phase transition around the black hole can arise.
The QCD phase transition \cite{Cline:1996mk}
and the electroweak (EW) phase transition
\cite{Nagatani:1998gv, Nagatani:2001nz}
have been discussed.
The local phase-transition around the black hole
means the formation of the spherical domain wall
which separates region of the symmetric-phase and that of the broken-phase.
The baryon number production and the electroweak baryogenesis scenario
have been proposed as an application of the spherical wall
\cite{Nagatani:1998gv, Nagatani:2001nz, Nagatani:1998rt}.

The Hawking temperature should be much greater than
the critical temperature of the phase transition
to consider the formation of the wall with the thermal phase transition.
The mean free paths for the radiated particles are much longer than
the Schwarzschild radius
and
we should confirm the local thermal equilibrium around the black hole
to consider the thermal phase transition.
Therefore
the Hawking temperature should be much higher than the critical temperature
to guarantee the local thermal equilibrium with the critical temperature.
However, it is natural to ask if
the Hawking radiation whose temperature is
equal to or greater than the critical temperature influences
the Higgs vacuum expectation value (vev) locally.

In this paper,
we discuss a Hawking radiation in the gauge-Higgs-Yukawa systems,
e.g., the Standard Model (SM) and the Grand Unified Theory (GUT),
and we show that
a structure of the spherical domain wall around the black hole
arises by the field dynamics of the system.
It seems that we need to consider the full dynamics of the field theory
to discuss the phenomenon around the black hole
and it will be technically difficult,
however,
only a part of interaction is relevant to the subject
because the mean free paths for the gauge interaction
is much longer than the radius of the black hole
and it simplifies the analysis.
The relevants are interactions
between the Hawking-radiated particles and the Higgs vev.
The radiated particles can be regard as ballistic
because of their long mean-free-paths.
By founding on these properties,
we propose a effective model with a {\it ballistic approximation}.
Our model consists from
the action for the Higgs field and
that for the relativistic point particles radiated from the black hole.
We derive an effective action for the Higgs field
which determines the structure of the Higgs vev around the black hole.
By considering the effective action,
the formation of the spherical domain wall around the black hole
is shown when the Hawking temperature is
equal or greater than the energy scale of the Higgs-Yukawa system.
We will call this mechanism
the {\it dynamical formation of the spherical wall}
in distinction from the thermal formation of the spherical wall.
As realistic cases,
the formation of the EW wall and
that of the GUT wall are discussed.

We also discuss a mechanism of 
the {\it spontaneous charging-up of the black hole}
as one of applications of the spherical wall.
We assume the following two conditions:
(i) a CP-broken phase in the spherical wall and
(ii) a chiral charge assignment of fermions in the field theory,
namely, a C-violation of the theory.
The Hawking radiation is neutral for the charge,
however the spherical wall has a reflection-asymmetry for the charge
on the assumptions.
Therefore
a charge-transport-mechanism from the wall to the black hole works
and it charges-up the black hole.
This mechanism is a variant application of
the charge transport scenario in the electroweak baryogenesis model
proposed by Cohen, Kaplan and Nelson \cite{Cohen:1991it, Cohen:1993nk}.
The two assumptions are equivalent to two of the Sakharov's three conditions
for the baryogenesis \cite{Sakharov:1967dj}.
In the case of the black hole whose Hawking temperature is EW scale,
the black hole can obtain a large amount of the hyper charge.
The black hole with GUT temperature can also obtain several charge.


The paper is organized as follows;
In Section \ref{Formation.sec},
the formation mechanism of the spherical wall is discussed.
In Section \ref{ChargeUp.sec},
a mechanism for the spontaneous charging-up of the black hole
is discussed.
In section \ref{summary.sec}
we provide a conclusion and discussions.

\section{Dynamical Formation of Spherical Domain Wall}
\label{Formation.sec}

We will consider a Hawking radiation
in the gauge-Higgs-Yukawa system like the Standard Model (SM).
Especially,
we will discuss dynamics of the Higgs field
around the black hole whose Hawking temperature is
the same energy scale of the system, e.g., the critical temperature.
In the situation,
the mean free paths from the gauge interactions are much greater
than the length scale for the black hole \cite{Nagatani:1998gv},
therefore, most of the gauge interactions are not important.
The relevants are the Yukawa interactions
for the heavy fermions and
the gauge interactions between the Higgs and heavy gauge bosons.
These particles obtain their heavy mass
from the Higgs vacuum expectation value (vev) through the interactions.
Inversely we can expect
that the particles radiated from the black hole
with high energy and with high density deform the Higgs vev
around the black hole.
It seems that we need full calculation for the field dynamics
to consider the deformation of the Higgs vev exactly.
However
we can employ the {\it ballistic approximation} for the radiated particles
because the mean free paths are enough long and
the only yukawa-type interactions are relevant.
Namely, the deformation of the Higgs vev can be discussed by
the relativistic semi-classical kinematics of
the each radiated ballistic particles through the relevant interactions.
The approximated system should have Lorentz invariance
and any ballistic particles obtain their mass by the Higgs vev.
To discuss the deformation in the ballistic approximation,
we propose a kind of Higgs-Yukawa model with relativistic point particles
which satisfies the required conditions and
describes the relevant interactions efficiently.
Our system is described by the action for a Higgs field $\phi(x)$
with a Higgs potential $V(\phi)$
and the action for relativistic point particles $\{y_i^\mu(s)\}$.
We will use metric convention $g_{\mu\nu} = {\rm diag} (+1,-1,-1,-1)$
and the action is given by a Nambu-Goto type action\footnote{
The action for the ordinary relativistic point particle with mass $m$
is given by $-\int ds \: m \sqrt{|\dot{y(s)}|^2}$.
Our action describes point particles which obtain their mass 
by the Higgs vev.}:
\begin{eqnarray}
 S[\phi,y] &=&
  \int d^4x \left[ (\partial \phi)^2 - V(\phi) \right]
  \;-\;
  \sum_i \int ds_i \; Y_i \, |\phi\left(y_i(s_i)\right)| \,
         \sqrt{|\dot{y}_i(s_i)|^2},
  \label{action}
\end{eqnarray}
where $\dot{y}_i^\mu \equiv dy_i^\mu / ds_i $
and $Y_i$ is a Yukawa coupling constant for the point particle $i$.
The summation in \eq{action} takes over all particles.
The Higgs potential in the vacuum is given by the double-well form:
\begin{eqnarray}
 V(\phi) &=& -\mu^2 \phi^2 \;+\; \frac{\mu^2}{v^2} \, \phi^4,
\end{eqnarray}
which has a minimum at $|\phi| = v/\sqrt{2}$
and the constant $\mu^2 > 0$ is the Higgs mass,
therefore mass of the particle $i$ is given by
$m_i = Y_i \left<\phi\right> = Y_i v / \sqrt{2}$ in the vacuum.

To calculate effective action for the Higgs field,
we will fix gauge for the point particles as $s_i = x^0 = y_i^0(x^0)$.
Then we have
$\sqrt{|\dot{y}_i(x^0)|^2} = \sqrt{1 - |{\bf v}_i(x^0)|^2} \equiv
1/\gamma_i(x^0)$,
where ${\bf v}_i(x^0)$ is 3-velocity for the particle $i$.
The action becomes
\begin{eqnarray}
  S[\phi,{\bf y}] &=&
  \int d^4x
  \left[ (\partial \phi)^2 - V(\phi) 
  \;-\;
  \sum_i \; \delta^{(3)} ({\bf x} - {\bf y}_i(x^0)) 
  Y_i \, \frac{|\phi\left(x\right)|}{\gamma_i(x^0)}
  \right]
  \label{action2}.
\end{eqnarray}
By using the definition of the 4-momentum for the particle $i$:
\begin{eqnarray}
 p_{i\,\mu}
  &=& \frac{\partial L}{\partial \dot{y}_i^\mu}
  \;=\; - Y_i \: \phi(y_i) \frac{\dot{y}_{i\,\mu}}{\sqrt{|\dot{y_i}|^2}}
  \;=\; - Y_i \: \phi(y_i) \gamma_i \; (1, \, -{\bf v}_i)
  \;\equiv\; (-E_i, \, {\bf p}_i),
\end{eqnarray}
the equation of motion for the particle becomes
\begin{eqnarray}
 \dot{p}_{i\mu} \;+\; Y_i \frac{1}{\gamma_i} \partial_\mu \phi &=& 0
  \label{particle_eom}.
\end{eqnarray}
The propagating modes for the Higgs can be regarded as
being taken into the ballistic particles
due to the ballistic approximation,
therefore the Higgs field $\phi(x)$ can be considered as
the Higgs expectation value and it is independent of time $x^0$,
namely, $\partial_0 \phi = 0$.
This assumption simplifies the equation of the motion in \eq{particle_eom}
and
\begin{eqnarray}
 E_i &=& Y_i \phi \gamma_i \label{Econst}
\end{eqnarray}
becomes a constant of the motion.
The constant $E_i$
is the energy of the particle $i$ at the rest frame.
When we put the trajectories for all particles $\{{\bf y}_i(x^0)\}$
with their energy-constants $\{E_i\}$,
we obtain the effective action for the Higgs field
\begin{eqnarray}
  S_\eff[\phi] &=&
  \int d^4x
  \left[ (\partial \phi)^2 - V(\phi) 
  \;-\;
  \sum_i \; \delta^{(3)} ({\bf x} - {\bf y}_i(x^0)) 
  Y_i^2 \, \frac{\phi^2(x)}{E_i}
  \right].
  \label{action3}
\end{eqnarray}
Here we adopt the differential particle number-density
$dE \times \N_{f}(x;E)$ for particle species $f$.
The effective action for the Higgs field can be written down as
\begin{eqnarray}
  S_\eff[\phi] &=&
  \int d^4x
  \left[ (\partial \phi)^2 - V(\phi) 
  \;-\;
  \phi^2
  \sum_f Y_f^2
  \int \frac{dE}{E} \N_f(E;x)
  \right],
  \label{action4}
\end{eqnarray}
where the summation in the \eq{action4} takes over
all particle-species in the theory.
Then the effective potential should be
\begin{eqnarray}
 V_\eff(\phi, x) &=& +\mu_\eff^2(x) \phi^2 \;+\; \frac{\mu^2}{v^2} \, \phi^4,
\end{eqnarray}
where we have defined the effective $\mu^2$ as
\begin{eqnarray}
 \mu_\eff^2(x)
  &=& -\mu^2
  \;+\; \sum_f Y_f^2 \int \frac{dE}{E} \N_f(E;x).
  \label{mueff}
\end{eqnarray}

Next we consider that the particle distribution $\N_f(E;x)$
is produced by the Hawking radiation form the Schwarzschild black hole
with mass $m_\BH$.
By the Hawking process,
the black hole radiates all particles in the field theory
with Hawking temperature $T_\BH =
\frac{1}{8\pi}\frac{m_\planck^2}{m_\BH}$
from the horizon whose radius is given by the Schwarzschild radius
$r_\BH = \frac{1}{4\pi}\frac{1}{T_\BH}$.
The particle distribution near the black hole is
approximately given by
\begin{eqnarray}
 dE \; \N_f
  &=& \frac{1}{4} \frac{g_f}{(2\pi)^3} \; f_{T_\BH}(E)
  \; 4\pi E^2 dE  \;\times\; \left(\frac{r_\BH}{r}\right)^2,
  \label{distribution}
\end{eqnarray}
where
\begin{eqnarray}
 f_{T_\BH}(E) &:=& \frac{1}{e^{E/T_\BH} \pm 1}
\end{eqnarray}
is Bose-Einstein or Fermi-Dirac distribution function
with temperature $T_\BH$.
The leading contribution from the Hawking radiation
is given by the \eq{distribution}.
We have ignored the backreaction from the produced wall into the
distribution of the particles,
namely,
we have ignored the contribution from the
non-uniform configuration of the Higgs field
which is depending on the distance from the black hole.
We note that
the ignored contribution to the particle trajectories
can be evaluated by solving the equation
$|\dot{\bf y}_i|^2 = 1 - (Y_i \phi({\bf y}_i)/E_i)^2$
derived from \eq{Econst}
and it will be a future subject.

\begin{figure}
 \begin{center}%
  \includegraphics{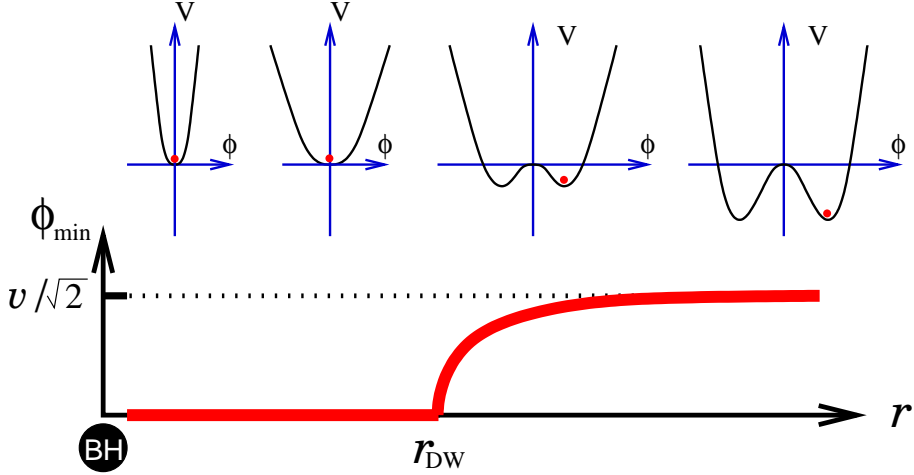}%
 \end{center}%
 \caption{%
 The distribution of the effective Higgs potential $V_\eff(\phi,r)$
 around a black hole.
 The parameter $r$ means the distance from the center of the black hole
 and $r_\DW=\alpha/\mu$ is the radius
 where the sign of $\mu_\eff(r)$ is inverted.
 The thick curves indicates the value of the Higgs field $\phi_\mini(r)$
 which minimizes the effective Higgs potential at the each point $r$.
 The value
 $v/\sqrt{2}$ describes the ordinary Higgs vacuum expectation value,
 namely $\phi_\mini(r\rightarrow\infty) = v/\sqrt{2}$.
 \label{WallConcept.eps}%
 }%
\end{figure}

By substituting the \eq{distribution} into the \eq{mueff},
we obtain
\begin{eqnarray}
 \mu_\eff^2(r)
  &=& -\mu^2
  \;+\; \frac{\alpha^2}{r^2},
\end{eqnarray}
where 
\begin{eqnarray}
 \alpha^2 &\equiv& \frac{1}{768\pi} \sum_f Y_f^2 \tilde{g}_f
  \label{alpha}
\end{eqnarray}
is a constant depending on the field theory
and we have defined the effective $g_f$ as
\begin{eqnarray}
 \tilde{g}_f &=&
  \left\{
   \begin{array}{cl}
    g_f & (f:{\rm boson})\\
    \frac{1}{2}g_f & (f:{\rm fermion})
   \end{array}
  \right.
\end{eqnarray}
Here we find that
the sign of $\mu_\eff(r)$ is reversed at $r_\DW \equiv \alpha/\mu$,
namely,
$\mu_\eff(r)$ is negative for $r>r_\DW$ and
$\mu_\eff(r)$ is positive for $r<r_\DW$.
Therefore,
the local-vacuum for the Higgs field is depending of the distance form the
black hole and we can expect the existence of
the spherical domain wall with radius $r_\DW$
when $r_\BH < r_\DW$.
The distribution of the Higgs potential $V(\phi,r)$ around the black hole
is shown in \fig{WallConcept.eps}.
The value of the Higgs field which minimizes the effective potential is
\begin{eqnarray}
 \phi_\mini(r) &=&
  \left\{
   \begin{array}{c@{\hspace{8mm}}l}
    \displaystyle
     \frac{v}{\sqrt{2}}
     \left[ \; {1 - \left(\frac{r_\DW}{r}\right)^2} \; \right]^{1/2}
     & (r > r_\DW) \\[5mm]
     0 & (r \leq r_\DW)
   \end{array}
  \right..
\end{eqnarray}
The (continuous varying) vacuum depending of the position
is one of the distinctive feature in the non-equilibrium system.
Such situations and the relation to the formation of the domain wall
had been discussed by several authors \cite{Cline:1996mk, Nagatani:1998gv}.

\begin{figure}[htbp]%
 \includegraphics{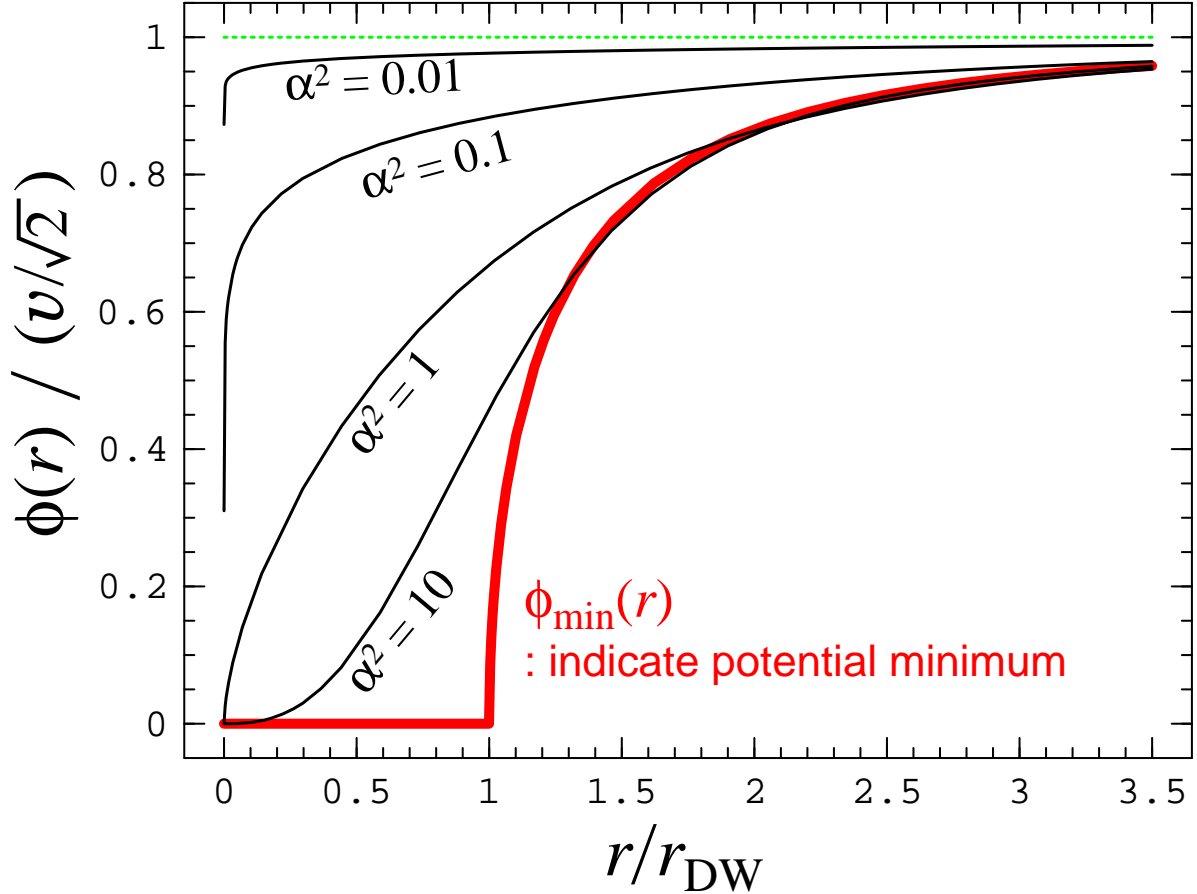}%
 \caption{%
 The profiles of the spherical-wall around a black hole.
 The parameter $r$ means distance from the center of the black hole
 and $\phi(r)$ means the expectation value for the Higgs field around the 
 black hole.
 The thin curves are stationary solutions of the EOM for the Higgs field
 with the effective potential $V_\eff(\phi,r)$
 depending on the parameter $\alpha^2$.
 The thick curve indicates the value of the Higgs field $\phi_\mini(r)$
 which gives the minimum of the effective Higgs potential
 and the dotted line means the ordinary Higgs vev
 $\left<\phi\right> = v/\sqrt{2}$.
 \label{WallFrom.eps}%
 }%
\end{figure}

The form of the Higgs vev around a black hole,
namely, the structure of the domain wall
can be calculated as a spherically-symmetric stationary solution of the
motion equation of the Higgs field with the effective potential:
\begin{eqnarray}
 -\Delta\phi(r) = -\frac{\partial}{\partial\phi} V_\eff(\phi,r),
  \label{higgseom}
\end{eqnarray}
where the boundary condition $\phi(r\rightarrow\infty) = v/\sqrt{2}$
is required.
The \eq{higgseom} can be solved numerically and
the result is shown in the \fig{WallFrom.eps}.
The form of the Higgs vev is depending on the parameter $\alpha$
which is defined in \eq{alpha}.
Generally speaking,
the curve of the Higgs vev $\phi(r)$ approaches the $\phi_\mini(r)$
when the parameter $\alpha$ becomes large.
Here we find that
the Hawking radiation can form the spherical wall
or wall-like structure around the black hole for the non-zero $\alpha$
and the radius of the spherical wall is characterized
by the parameter $r_\DW$.

The characteristic radius for the wall is given by $r_\DW = \alpha/\mu$
as discussed previously.
It should be noticed that the characteristic radius $r_\DW$
is not depending on the Hawking temperature
then the form of the wall is not so changing
after the formation of the wall to the evaporation of the black hole.
The Schwarzschild radius of the black hole $r_\BH$
should be smaller than the characteristic radius of the wall $r_\DW$
for the formation of the wall
then the condition for the wall-existence is
\begin{eqnarray}
 1 &<& \frac{r_\DW}{r_\BH} \;=\; 4\pi \alpha \frac{T_\BH}{\mu}.
  \label{condition}
\end{eqnarray}

Now we can discuss the formation of the domain wall
in the realistic field theories, e.g., the electroweak (EW) theory and
the grand unified theory (GUT).
In the case of the EW domain wall,
the heavy particles which contribute to create the wall
are the top quarks, the weak bosons $Z^0$, $W^{\pm}$ and the Higgs bosons,
and approximately we have
\begin{eqnarray}
 \alpha_\EW^2 &\equiv& \frac{1}{768\pi} \sum_f Y_f^2 \tilde{g}_f
  \;\simeq\; 1/200.
  \label{alphaEW}
\end{eqnarray}
The critical Hawking-temperature for the dynamical formation of
the spherical EW wall becomes
\begin{eqnarray}
 T_\BH^{*\EW} &\equiv& \frac{\mu}{4\pi\alpha_\EW}
  \;\sim\; \mu.
\end{eqnarray}
Therefore
the Hawking radiation produces the spherical EW domain wall
whose structure is given in the \fig{WallFrom.eps}
with $\alpha^2 = \alpha_\EW^2 \simeq 0.005$
when the Hawking temperature of the black hole
is similar to or greater than the EW scale.
The characteristic radius of the wall is given by
\begin{eqnarray}
 r_\DW^\EW &=& \alpha_\EW/\mu \;\simeq\; 0.07/\mu
\end{eqnarray}
In the case of the $SU(5)$-GUT,
there are, at least, lepto-quarks $X,Y$ and $\bf 24$-Higgs $\Phi$
as heavy particles with the GUT scale mass $\mu_\GUT \simeq 10^{16}\GeV$.
The same analysis results
\begin{eqnarray}
 \alpha_\GUT^2 &\simeq& 1/70,\\
 T_\BH^{*\GUT} &\simeq& 0.7\mu_\GUT,\\
 r_\DW^\GUT    &\simeq& 0.1/\mu_\GUT.
\end{eqnarray}
Then the spherical GUT wall around the black hole is also formed
when the Hawking temperature is greater than the GUT scale.
Finally we can discuss that
the dynamical spherical wall around a black hole may be formed
in most of the field theory which has a Higgs mechanism
when the Hawking temperature is greater
than the energy scale of the field theory.

\section{Spontaneous Charging-up of the Black Hole}
\label{ChargeUp.sec}

We will discuss that the Hawking radiation can charge up the black hole
by the effect of the spherical domain wall which has been discussed.
We need two assumptions such that
(i) the domain wall has CP-broken phase and
(ii) the field theory has fermions with chiral charge assignment.
The Standard Model satisfies the second assumptions
because the left-handed quark and the right-handed quark have 
different hyper charges.
When the first assumption is satisfied,
the reflection rate on the wall of the left-handed fermions
is different from that of the right-handed fermions.
Therefore
the domain wall has a charge-reflection-asymmetry
when both assumptions are satisfied.
The Hawking radiation is charge-neutral, however,
the black hole obtains net charge by the effect of the domain wall
because a part of the reflected particles return into the black hole,
namely the net charge is transported to the black hole.
This process is similar to the
``charge transport mechanism/scenario by the thin wall''
in the electroweak baryogenesis
proposed by Cohen, Kaplan and Nelson \cite{Cohen:1991it, Cohen:1993nk}.
In the charge transport scenario of the electroweak baryogenesis,
the hyper charge is transported
from the thin EW wall to the region of the symmetric phase
and it boosts up the baryon number creation
by the sphaleron process.
On the other hands,
our charge-transportation charges up the black hole.

The charging-up-rate for the black hole is given by
\begin{eqnarray}
 \frac{dQ}{dt} = \sigma_\BH \times C_\focus F_Q,
\end{eqnarray}
where $\sigma_\BH$ is the cross section for the absorption to the black hole,
$F_Q$ is the reflected charge flux at the wall and
the dimensionless parameter $C_\focus$ is a focusing factor.
The absorption-cross-section for the Schwarzschild black hole is
given by
\begin{eqnarray}
  \sigma_\BH &=& \left\{
    \begin{array}{ll}
     \displaystyle 4\pi r_\BH^2 & (r_\BH \lnear r_\DW)\\
     \displaystyle \frac{27\pi}{4} r_\BH^2 & (r_\BH \ll r_\DW)\\
    \end{array}\right..
\end{eqnarray}
The cross section is given by the horizon area
when the radius of the black hole $r_\BH$
is similar to or a little smaller than 
the characteristic radius of the domain wall $r_\DW$.
In the case of $r_\BH \gg r_\DW$,
the reflected particles at the wall are regarded as
the incoming particles from the infinite distance to the black hole,
therefore we should adopt the absorption-cross-section as
$\sigma_\BH = \frac{27\pi}{4} r_\BH^2$ rather than the horizon area.
The difference among these cross sections is only factor one
and the spontaneous charging-up mechanism mainly works for
$r_\BH \lnear r_\DW$, then
we will use the horizon area as the cross section.

In the spherical reflector,
any particles radiated from neighborhood of the center of the reflector
return to neighborhood of the center
when they are reflected.
Then the reflected flux at the horizon is different
from the reflected flux at the spherical wall.
This is a focusing effect by the spherical reflector.
The effect increases the flux at the horizon
compared with the flux at the wall $F_Q$.
The flux at the horizon can be written down as $C_\focus F_Q$
by the focusing factor $C_\focus \geq 1$.
We have $(r_\BH/r_\DW)^2 \times C_\focus = 1$ when the effect maximally works
and we have $C_\focus = 1$ in the absence of the effect.

The charge flux at the wall is given by
\begin{eqnarray}
 F_Q &=& \sum_{f \in {\rm Fermions}} \;
  \int_{E > m_f} dE \; {\cal N}_f(E ; r_\DW) \;
  \Delta Q_f \;
  \Delta R_f(E).
  \label{QFlux}
\end{eqnarray}
The summation in the equation
is taking over all species of the chiral-charged fermions
and the particle-species $f$ does not distinguish both
the chirality of the particle and the particle/anti-particle.
The number $\Delta Q_f \equiv Q_{f_L} - Q_{f_R}$
means the difference 
between the charge of left-handed fermion $f_L$ and
that of right-handed fermion $f_R$.
These numbers are related to the C-violation for the theory.
The value
\begin{eqnarray}
 \Delta R_f(E) &\equiv&
  R_{f_R \rightarrow f_L}(E) - R_{\bar{f}_R \rightarrow \bar{f}_L}(E)
\end{eqnarray}
describes the difference of the reflection-probabilities,
where
the reflection-probability $R_{f_R \rightarrow f_L}(E)$ means 
a probability of that the left-handed fermion $f_R$ with energy $E$
is reflected to the right-handed fermion $f_L$
and
the reflection-probability $R_{\bar{f}_R \rightarrow \bar{f}_L}(E)$ means
the same for the anti-fermions $\bar{f}_R \rightarrow \bar{f}_L$.
These probabilities are functions for the energy of the particles $E$.
The non-zero value of $\Delta R_f(E)$
is related with the CP-broken phase assumed in the wall
and can be calculated in the way discussed by the Cohen Kaplan and Nelson
\cite{Cohen:1991it, Cohen:1993nk}
when we put the profile of the CP-broken phase in the wall.

\begin{figure}[htbp]%
 \includegraphics{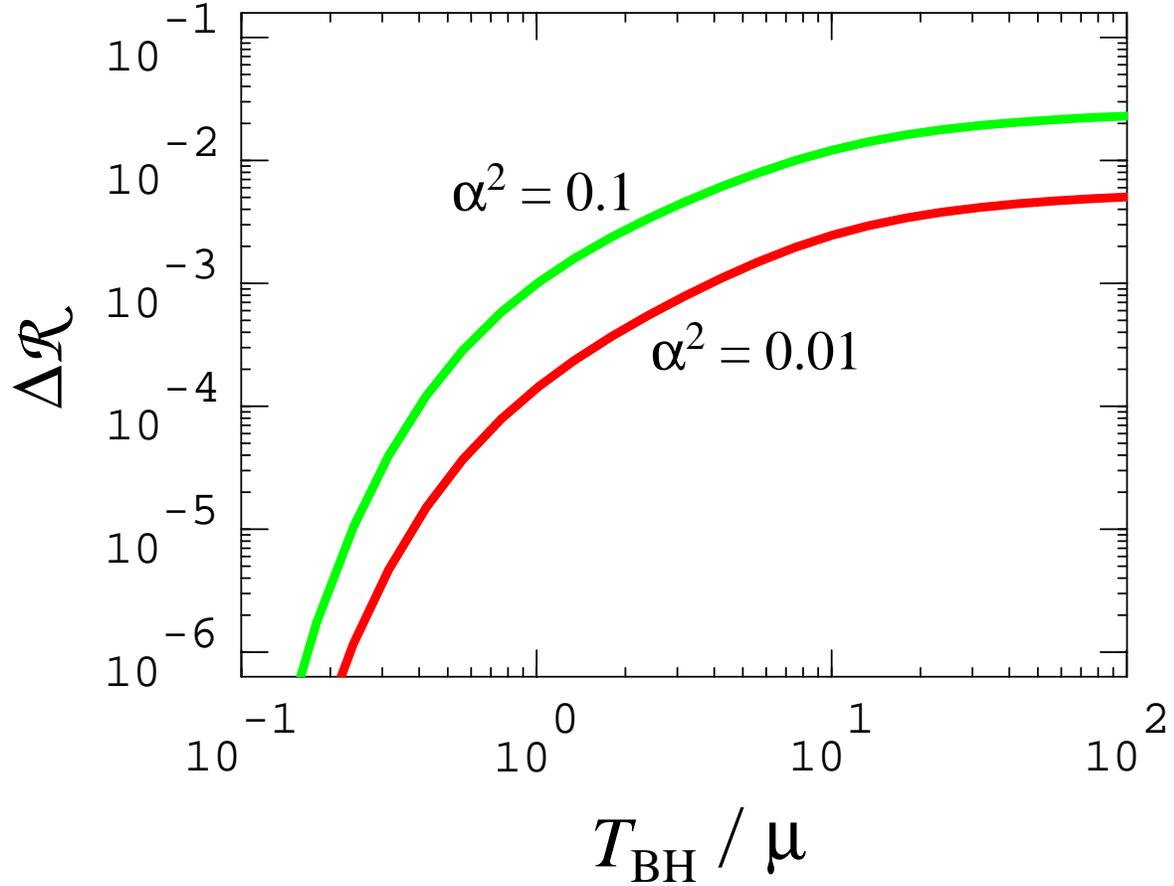}%
 \caption{%
 Numerical results of the 
 dimension less reflection asymmetry $\Delta{\cal R}(m_f,T_\BH)$
 for the wall given in \eq{CPprofile} with
 $\alpha^2=0.1$ and $\alpha^2=0.01$.
 We have assumed $\mu = m_f$ for simplicity.
 \label{DR.eps}%
 }%
\end{figure}

To evaluate the reflection asymmetry $\Delta R_f(E)$
of the wall around the black hole,
we assume the profile of the CP-broken phase in the wall:
\begin{eqnarray}
 \left<\phi(r)\right>
  &=&
  f(r) \times \exp
  \left[
   \Delta\theta_\CP \left( 1 - \frac{f(r)}{v/\sqrt{2}} \right)
  \right], \label{CPprofile}
\end{eqnarray}
where $\Delta\theta_\CP$ is the amount of the CP-broken phase in the wall
and the function $f(r)$ is the profile of the Higgs field in the wall.
The profile function $f(r)$ is
solution of the motion equation of the Higgs field with effective
potential in the \eq{higgseom} and
it is depending of the parameter $\alpha$ defined in \eq{alpha}.
The wall given in \eq{CPprofile} is defined for $r>r_\BH$.
By substituting \eq{distribution} into \eq{QFlux},
the charge flux becomes
\begin{eqnarray}
 F_Q
  &=&
  \left(\frac{r_\BH}{r_\DW}\right)^2 C_\focus
  \,\times\,
  \sum_{f} \;
  g_f \, \Delta Q_f \, m_f^3 \;
  \Delta{\cal R}(m_f,T_\BH),
\end{eqnarray}
where we have defined dimensionless function
depending on the wall profile
\begin{eqnarray}
 \Delta{\cal R}(m_f,T_\BH)
  &=&
  \frac{1}{8\pi^2}
  \int_1^\infty d\xi \;
  \xi^2 \; f_{T_\BH}(m_f \xi) \;
  \Delta R_f(m_f \xi).
\end{eqnarray}
We have numerically evaluated $\Delta R_f(E)$ and
$\Delta {\cal R}_f(m_f,T_\BH)$
for the wall in \eq{CPprofile}
according to the method proposed by Cohen Kaplan and Nelson
\cite{Cohen:1991it}.
In the case of maximum CP-violated wall $\Delta\theta_\CP = \pi$,
we obtain the dependence of the dimensionless coefficient
$\Delta {\cal R}_f(m_f,T_\BH)$ in \fig{DR.eps}.
In this calculation we have assumed $\mu = m_f$ for the simplicity.

The black hole with the Hawking temperature $T_\BH$
has a finite lifetime
$\tau_\BH = \frac{20}{\pi^2 g_*} \frac{m_\planck^2}{T_\BH^3}$,
where $g_*$ is the total degree of freedom with a fermion correction.
The total charge transported to the black hole in his lifetime
by the effect of the wall is
\begin{eqnarray}
 Q
 &=&
  \int^{\tau_\BH} dt \frac{dQ}{dt} \nonumber\\
 &=&
  \frac{15}{\pi^3 g_*} m_\planck^2
  \sum_f g_f \Delta Q_f m_f^3
  \int_{T_\BH}^\infty \frac{dT}{T^6}
  \Delta{\cal R}(m_f,T), \label{charge}
\end{eqnarray}
where we have assumed the maximum focusing effect.
The integration in the \eq{charge} can be performed
by using the numerical form of $\Delta{\cal R}$.
The numerical results $\Delta{\cal R}$ with $\mu = m_f$ in \fig{DR.eps}
have meaningful value for $T_\BH>\mu$ and
they are exponentially dumping for $T_\BH<\mu$.
Therefore we have approximately
\begin{eqnarray}
 Q
 &\simeq&
  \frac{3}{\pi^3} \,
  \Delta{\cal R}(\mu,\mu)
  \left(\frac{m_\planck}{\mu}\right)^2
  \sum_f
  \frac{g_f}{g_*}
  \Delta Q_f,
\end{eqnarray}
where we have assumed a simplification $m_f = \mu$
for all related fermions $f$
and we have assumed the initial Hawking temperature $T_\BH$
is similar to or smaller than the energy scale of the field theory $\mu$.
Our numerical analysis results
$\Delta{\cal R}(\mu,\mu) \simeq 10^{-4}$ for $\alpha^2=1/100$
as the realistic field theories,
then we have
\begin{eqnarray}
 Q &\simeq& 
  10^{-5}
  \times
  \left(\frac{m_\planck}{\mu}\right)^2
  \sum_f \frac{g_f \Delta Q_f }{g_*}.
\end{eqnarray}
Finally we conclude that
the spherical wall can charge up the black hole non-trivially
when the energy scale of the wall satisfies
$\mu \lnear 10^{-3} \times m_\planck \simeq 10^{16}\GeV$.

In the case of a black hole whose Hawking temperature is
the EW scale $\mu\simeq100\GeV$,
the spherical EW wall arises and
the hyper charge $Y$ is transported to the black hole
by assuming CP-broken phase in the wall.
Mainly the top quarks, $g_f = 3$ and $\Delta Q_f = 1/2$,
carry the charge to the black hole.
The total transported hyper charge is given by $Q \simeq 10^{27}$,
then we consider the spontaneous charging up mechanism
of the black hole can work strongly.
In the case of a black hole with GUT temperature,
several charge can be transported to the black hole.
The amount of the charge is
depending on the chiral charge assignment $\Delta Q_f$ of the GUT.

\section{CONCLUSION AND DISCUSSIONS}\label{summary.sec}

%
In this paper
the Hawking radiation in a kind of Higgs-Yukawa system is discussed
and a dynamical formation of the spherical domain wall is shown.
An action for the many relativistic point particles
(Nambu-Goto like action) with Higgs coupling and
the ordinary Higgs action with double well potential are
adopted to describe the system.
We expect that the system approximately describes
the Higgs-Yukawa system
which is consist from the ordinary Dirac action and the bosonic actions.
Our action simplifies the analysis the Higgs vev structure
around a black hole as compared with solving the Dirac/Weyl equations.

The black hole has been assumed as an heat source
with Hawking temperature in our analysis,
namely,
the general relativistic effects are omitted.
Reliable treatments for general relativistic corrections
for the Hawking-radiated particles
do not have been known.
For example, the blue-shift effect near the horizon
for the radiated particles is arise
when we consider that
the radiated particles obeys the Schwarzschild metric.
The blue-shift effect implies that
the energy of the radiated particles near horizon is
much higher than the Hawking temperature
and we can discuss that any low Hawking-temperature black hole
produces the domain wall dynamically or thermally,
therefore, the effect is not acceptable widely.

In our calculation,
we have evaluated the influence from the radiated particles
into the effective potential for the Higgs field,
however,
the influence from the wall into the motion of the radiated particle,
namely, the backreaction is neglected for simplicity.
The form of the spherical wall may be slightly deformed by 
the backreaction.
The issue of the backreaction will be future subject.

We have also discussed the mechanism of the spontaneous charging up
of the black hole by the effect of the spherical domain wall.
The mechanism can work
when C-violation of the field theory and
CP-broken phase in the wall are assumed.
Our analysis results that the mechanism can work
when the initial Hawking temperature of the black hole
is smaller than about $10^{16}\GeV$.
The spherical EW domain wall by the black hole with EW temperature
can transport a large amount of hyper charge to the black hole.
The black hole with GUT temperature can obtain several charge.

We have discussed the spontaneous charging up of the black hole.
On the other hand,
a mechanism for the charge loss of the black hole
has been discussed.
Gibbons first proposed this subject
by the semi-classical method \cite{Gibbons:1975kk, Hiscock:1990ex}
and Gabriel discussed that this result can be confirmed recently
by the functional method \cite{Gabriel:2000mg}.
They discussed that 
the charged black hole loses its charge
by the pair creation of the charged particles
because of the strong electric field around the black hole,
namely, a kind of Schwinger process works to discharge the black hole.
These calculation are reliable only for $r_\BH > 1/m_e$
where $m_e$ is mass of electron,
namely the radius of the black hole should be greater
than the Compton wave length of the electron.
Several authors discuss the subject
for the charge loss of the smaller black hole $r_\BH < 1/m_e$
\cite{Page:1977um, Khriplovich:1998si},
however we do not have common understanding for this subject.
We have discussed the spontaneous charging-up of the black hole
whose radius is smaller than the EW scale,
therefore we can not apply these results of the charge-loss
directly to our system.
This subject may be related to the remnant of the black hole.
Zel'dovich discussed that
a black hole leaves a remnant with a Planck mass scale
after the end of its Hawking radiation \cite{Zeldovich:1984}.
Both our mechanism for spontaneous charging-up
and some mechanism for the charge-loss may be working
at the final stage of the black hole evaporation.
We may expect that
a remnant with several (hyper) charge
will be left after the end of the Hawking radiation.

\begin{flushleft}
 {\Large\bf ACKNOWLEDGMENTS}
\end{flushleft}

 We would like to thank
 M.~Nojiri, M.~Ninomiya, K.~Kohri and K.~Shigetomi
 for their useful discussions and suggestions.
 YN is indebted to the Japan Society for
 the Promotion of Science (JSPS) for its financial support.
 The work is supported in part by a Grant-in-Aid for Scientific Research
 from the Ministry of Education, Culture, Sports, Science and Technology
 (No. 199903665).

\newcommand{\PRL}[3]	{{Phys.\ Rev.\ Lett.}   {\bf #1}, #2 (#3)}
\newcommand{\PR}[3]	{{Phys.\ Rev.}          {\bf #1}, #2 (#3)}
\newcommand{\PRA}[3]	{{Phys.\ Rev.\ A}       {\bf #1}, #2 (#3)}
\newcommand{\PRD}[3]	{{Phys.\ Rev.\ D}       {\bf #1}, #2 (#3)}
\newcommand{\PL}[3]	{{Phys.\ Lett.}         {\bf #1}, #2 (#3)}
\newcommand{\PLA}[3]	{{Phys.\ Lett.\ A}      {\bf #1}, #2 (#3)}
\newcommand{\PLB}[3]	{{Phys.\ Lett.\ B}      {\bf #1}, #2 (#3)}
\newcommand{\NuP}[3]	{{Nucl.\ Phys.}         {\bf #1}, #2 (#3)}
\newcommand{\PTP}[3]	{{Prog.\ Theor.\ Phys.} {\bf #1}, #2 (#3)}
\newcommand{\Nature}[3]	{{Nature}               {\bf #1}, #2 (#3)}
\newcommand{\PKNAW}[3]	{{Proc.\ K.\ Ned.\ Akad.\ Wet.} {\bf #1}, #2 (#3)}
\newcommand{\Physica}[3]{{Physica\ (Utrecht)}   {\bf #1}, #2 (#3)}
\newcommand{\JMP}[3]	{{J.\ Math.\ Phys.}     {\bf #1}, #2 (#3)}
\newcommand{\PRSLA}[3]	{{Proc.\ R.\ Soc.\ London,\ Ser A} {\bf #1}, #2 (#3)}
\newcommand{\AP}[3]	{{Ann.\ Phys.\ (N.Y.)}  {\bf #1}, #2 (#3)}
\newcommand{\JPA}[3]	{{J.\ Phys.\ A}         {\bf #1}, #2 (#3)}
\newcommand{\ZhETF}[3]	{{Zh.\ \'{E}ksp.\ Teor.\ Fiz.\ Pis'ma.\ Red.} {\bf #1}, #2 (#3)}
\newcommand{\JETP}[3]	{{JETP\ Lett.}          {\bf #1}, #2 (#3)}
\newcommand{\CMP}[3]	{{Commun.\ Math.\ Phys.}{\bf #1}, #2 (#3)}
\newcommand{\MPLA}[3]   {{Mod.\ Phys.\ Lett.}   {\bf A#1}, #2 (#3)} 


\end{document}